\documentclass[11pt]{article}
\pdfoutput=1

\usepackage{setspace}
\usepackage{cite}

\renewcommand{\Phi}{\phi}

\usepackage[utf8]{inputenc}
\usepackage{geometry}
\usepackage[USenglish]{babel}
\usepackage{verbatim}
\usepackage{prettyref}
\usepackage{amsmath}
\usepackage[normalem]{ulem}
\usepackage{amssymb}
\usepackage{graphicx}
\usepackage{setspace}
\usepackage{esint}
\usepackage{verbatim}

\usepackage{color}
\definecolor{darkgreen}{rgb}{0,0.5,0}
\definecolor{darkblue}{rgb}{0,0,0.6}
\definecolor{purple}{rgb}{0.4,.2,0.7}
\definecolor{orange}{rgb}{0.95, 0.5, 0.3}

\usepackage[colorlinks=true,citecolor=darkgreen,linkcolor=darkblue,urlcolor=blue]{hyperref}

\numberwithin{equation}{section}
\numberwithin{table}{section}

\def\be{\begin{equation}}
\def\ee{\end{equation}}
\def\bea{\begin{eqnarray}}
\def\eea{\end{eqnarray}}
\def\ba{\begin{align}}
\def\ea{\end{align}}

\newcommand{\cE}{\mathcal{E}}

\makeatletter
\newcommand{\vast}{\bBigg@{4}}
\newcommand{\Vast}{\bBigg@{5}}
\makeatother

\begin{document}
\begin{spacing}{1.1}
  \setlength{\fboxsep}{3.5\fboxsep}

~
\vskip5mm

\begin{center} {\Large \bf Hydrodynamics without boosts}
%\begin{center} {\Large \bf An \ae thereal theory of hydrodynamics}

\vskip10mm

Igor Novak$^1$, Julian Sonner$^1$ \& Benjamin Withers$^{1,2}$\\
\vskip1em
{\small
{\it  $^1$Department of Theoretical Physics, University of Geneva, 24 quai Ernest-Ansermet, 1211 Gen\`eve 4, Suisse}
{\it  $^2$Mathematical Sciences and STAG Research Centre,University of Southampton, Highfield, Southampton SO17 1BJ, UK}
}
\vskip5mm

\tt{ igor.novak@unige.ch, julian.sonner@unige.ch, b.s.withers@soton.ac.uk}

\end{center}

\vskip10mm

\begin{abstract}

We construct the general first-order hydrodynamic theory invariant under time translations, the Euclidean group of spatial transformations and preserving particle number, that is with symmetry group  $\mathbb{R}_t\times$ISO$(d)\times$U$(1)$. Such theories are important in a number of distinct situations, ranging from the hydrodynamics of graphene to flocking behaviour and the coarse-grained motion of self-propelled organisms. Furthermore, given the generality of this construction, we are are able to deduce special cases with higher symmetry by taking the appropriate limits. In this way we write the complete first-order theory of Lifshitz-invariant hydrodynamics. Among other results we present a class of non-dissipative first order theories which preserve parity.

\end{abstract}

\pagebreak

\pagestyle{plain}

\setcounter{tocdepth}{2}
{}
\vfill
\tableofcontents
\section{Introduction}
Any system that finds itself in a state of local thermodynamic equilibrium, is thought to evolve to its global equilibrium state, described by universal long-wavelength, long time-scale hydrodynamics, respecting positivity of entropy production. The universal theory governing this dynamics, corresponds to the relaxation of all conserved quantities of a given system, and can be adapted to particular physical situations of interest by specifying an equation of state, as well as the functional form and value in equilibrium of a specified set of transport coefficients \cite{landau1987fluid}.

Hydrodynamics is an extremely successful practical example of the philosophy of effective field theory. Its equations are formulated by arranging terms in ascending order of derivatives, truncating at a specified order in this expansion\footnote{For a clear review of this procedure in a relativistic system, see \cite{Kovtun:2012rj}}. The possible terms that may appear in this expansion are restricted by the symmetries as well as the usual rules of effective field theories, in such a way as to reduce an a-priori redundant set of transport coefficients to a smaller, purely physical subset. The fewer symmetries one requires, the more general the resulting framework, and the larger the number of allowed transport coefficients. From this one may recover previously known, more symmetric versions by taking appropriate limits.

The main impetus to develop the general theory, apart from being structurally interesting, is of course that situations  arise in which the least symmetric theory is the only one applicable. For example, in order to clarify the distinction between ordinary quasi-normal modes and the spatial collective modes of \cite{Sonner:2017jcf,Novak:2018pnv} one should look to non-boost invariant systems. A non-relativistic hydrodynamic theory without Galilean invariance is furthermore needed to describe the electron fluid of graphene at finite carrier density \cite{doi:10.1002/andp.201700043,Lucas:2017idv}. 
Another area where such examples are relevant is biophysics. Consider, for example, the case of a system of a large number of {\it self-propelled} organisms such as birds, moving through a {\it fixed medium}, such as the air. A coarse grained description of such a collection of self-propelled ``particles'' in terms of fluid dynamics will be translation invariant, but not invariant under any form of boosts. For a perspective of applying non boost-invariant hydrodynamics to flocking behavior of birds, the reader may want to consult \cite{TONER2005170}. Applications to the theory of active matter are discussed in \cite{PhysRevLett.92.118101,callan2011hydrodynamics,furthauer2012taylor}.

In the spirit of proceeding from the most general to the more specific, in this paper we formulate the complete first-order theory of hydrodynamics invariant under time translations $\mathbb{R}_t$, the Euclidean group of spatial symmetries ISO$(d)$ and containing a conserved charge or particle number leading to the total symmetry group $\mathbb{R}_t\times$ISO$(d)\times$U$(1)$. Such a theory does not possess any form of boost symmetry, be it of non-relativistic Galilean or relativistic Lorentz form. We also explain how to recover previously known examples of Galilean-invariant, Lifshitz-invariant and relativistically invariant hydrodynamics as a limiting procedure. In all these the number of transport coefficients is reduced, sometimes drastically.

We approach this theory from the non-relativistic, non-boost invariant side, but it is also interesting and informative to ask the opposite question: if we were to start with a relativistically invariant theory, what different patterns of symmetry breaking and what kind of non-relativistic and non-boost invariant structures can possibly arise? This was answered for equilibrium configurations in \cite{Nicolis:2015sra}, where the resulting states can be classified according to eight different symmetry-breaking patterns, according to how the remaining generators of Poincar\'e are twisted with internal symmetries. In their language we are developing the hydrodynamics of a``type-I framid", which breaks full Poincar\'e invariance down to only translations and rotations without any internal symmetry twist, in other words to $\mathbb{R}_t\times$ISO$(d)\times G_\textrm{internal}$. We will exclusively focus on the case where the internal symmetry is U$(1)$. As was already noted in \cite{Nicolis:2015sra} this pattern of symmetry breaking is closely related to Einstein-Aether theory \cite{Jacobson:2000xp} and can be seen as the breaking of Poincar\'e invariance induced by a time-like expectation value of a vector operator.

The authors of \cite{deBoer:2017ing,deBoer:2017abi} analyse linearised hydrodynamic fluctuations at first order in the derivative expansion for fluid flows at rest with respect to the preferred reference frame. As our analysis in the following shows, the more general case of arbitrary velocity with respect to the preferred reference frame, forces one to introduce a larger number of additional transport coefficient and thus exhibits new physics associated with these.

\section{Non-boost-invariant hydrodynamics}
\begin{figure}[t]
\begin{center}
\includegraphics[width=\columnwidth]{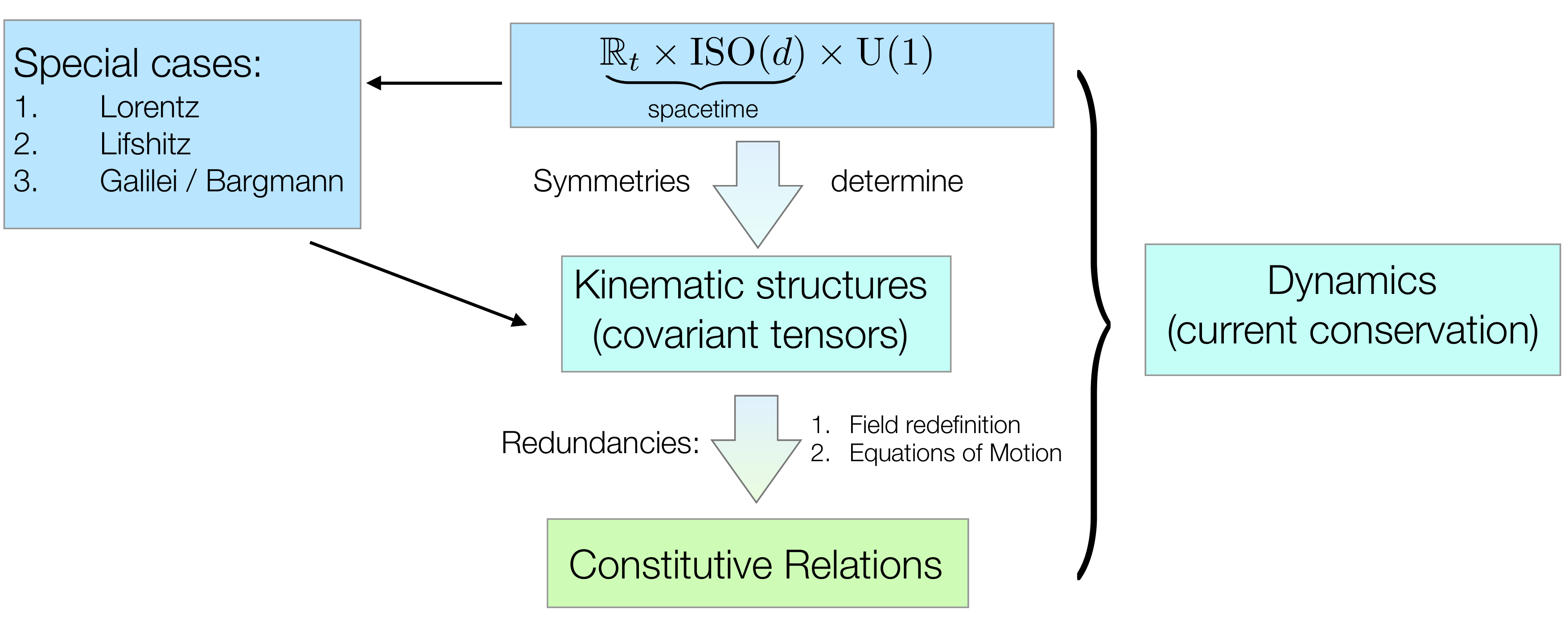}\\
\end{center}
\caption{Here we give the roadmap of our construction. In the spirit of effective field theory, we first identify the most general symmetry group applicable for our purposes. For the purpose of this paper we focus on $\mathbb{R}_t \times \textrm{ISO}(d) \times \textrm{U}(1)$. Next we write down all possible kinematic structures compatible with that symmetry, which are then constrained by the fixing redefinition ambiguities present in the construction. Finally the remaining physical quantities are then subject to a dynamical principle, which in the case of hydrodynamics follows simply from conservation of the stress tensor and current. From the most general and therefore least constrained structure we also recover more highly symmetric previously known examples as limiting cases as shown in the leftmost box.}
 \label{fig:TheMethod}
\end{figure}

Figure \ref{fig:TheMethod} gives a conceptual overview of the procedure we follow in order to construct the general non boost-invariant hydrodynamic theory at first dissipative order. The challenge in this construction lies in the large number of allowed tensor structures (45 in the general setting) and transport coefficients (29 in the general setting\footnote{This count includes both dissipative and non-dissipative transport coefficients at first order. Eliminating the latter class reduces the count further to 20.}) and so it is essential to ensure one includes all terms and to be calculationally as efficient as possible. Hydrodynamics, being defined as a gradient expansion, contains the usual field redefinition ambiguities inherent in effective field theory constructions. At first order -- the highest order to we explore in this work -- one may adjust the coefficients of a certain subset of tensor structures, by a) using the zeroth-order equations of motion and b) by redefining the zeroth-order hydrodynamic variables (temperature, velocity and chemical potential) by first-order shifts. This means that not all coefficients appearing at first order are physical transport coefficients and our goal becomes to isolate only those that are. Once all equations of motion have been imposed, the remaining ambiguities are those associated to shifts of the hydrodynamic variables. In a boost-invariant setting these ambiguities are typically fixed by a choice of hydrodynamic frame, such as Landau frame. Here, in the non boost-invariant setting, we demonstrate similarly that all remaining ambiguities are fixed by an appropriate frame choice. It is encouraging that the resulting theory, in all limiting cases, reduces to the previously known constructions with the right number of transport coefficients and tensor structures.

\subsection{The ideal fluid}
We being our construction of the general first-order hydrodynamics theory with symmetry $\mathbb{R}_t\times$ISO$(d)\times$U$(1)$ by writing the constitutive relations up to first order in derivatives. This task is facilitated by the work of \cite{deBoer:2017ing,deBoer:2017abi}, who wrote down, in the first instance, the constitutive relations for an {\it ideal} fluid in this symmetry class. These authors write the constitutive relations in the laboratory frame where the fluid has velocity $v^i$:
\begin{align}
T^{(0)}{}^{0}\!_{0}  &= -\mathcal{E}, \quad T^{(0)}{}^{0}\!_{j} = \rho v^j, \quad T^{(0)}{}^{i}\!_{0}  = -(\mathcal{E}+P)v^i, \quad T^{(0)}{}^{i}\!_{j}  = P \delta^{i}\!_{j} + \rho v^i v^j.\\
J^{(0)}{}^{0}&= n, \quad J^{(0)}{}^{i}= n v^{i} 
\end{align}
In these expressions ${\cal E}$ and $P$ are the energy density and pressure, $n$ is the particle density or charge density (depending on the chosen interpretation of the U$(1)$ symmetry), while $\rho$ is the ``kinetic mass density" \cite{deBoer:2017ing}.  This additional thermodynamic function is generically different from the mass density due to the absence of a boost symmetry and must therefore be included independently in all boost-non-invariant cases. In total we thus have the thermodynamic functions $\cE, P, n$ and $\rho$, which are arbitrary functions of the thermodynamic variables, namely the temperature $T(t,x^i)$, the square of the velocity field $v^2(t,x^i)$ and the chemical potential $\mu(t,x^i)$. A related quantity that we will sometimes find convenient is the internal energy $\tilde\cE : = \cE - \rho v^2$. The principal interest of this paper is to extend this to the dissipative level, more precisely to first order in the hydrodynamic expansion, keeping arbitrary the velocity with respect to the preferred reference frame, $v^i$. A linear analysis in $v^i$ around $v^i = 0$ is performed in \cite{deBoer:2017abi}, giving the hydrodynamic modes in the preferred frame.

\subsection{Dissipative corrections}
The goal of this section is to write down the general constitutive relations for stress tensor and current at first order in derivatives with respect to the hydrodynamical variables.  Keeping with common notation in the literature we write
\bea
T^{\mu}{}_\nu &=& T^{(0)}{}^\mu{}_\nu + \Pi^\mu{}_\nu,\\
J^{\mu} &=& J^{(0)}{}^\mu + \Pi^\mu\,,
\eea
where $\Pi^\mu{}_\nu$ and $\Pi^\mu$ contain terms of first-order or higher in derivatives of the hydrodynamic variables. These contain the dissipative terms (in addition to several non-dissipative transport coefficients which we will discuss in detail).

As has been outlined above, there is a large degree of redundancy to be fixed, which stems from the way hydrodynamics is arranged as an expansion in derivatives. The ambiguity is usually fixed by making use of a particular `frame' (e.g. Landau frame). We will ultimately make such a choice as well, but not before systematically exploring precisely what ambiguity is present in the general non boost-invariant hydrodynamic theory, and that our choice of frame consistently fixes the ambiguity. A loose, but helpful, analogy here is the construction of a putative new gauge theory, where one would be interested to demonstrate that a particular gauge condition actually fixes the full redundancy.

Let us therefore now discuss in more detail what sort of redundancies arise in the construction.
\subsubsection{Field redefinitions}
The first class of redundancy we need to take into account comes from the fact that shifts of the form
\begin{align}
T(t,x^i) & \longrightarrow T(t,x^i) + \delta T(t,x^i), \nonumber \\
v^i(t,x^i) & \longrightarrow v^i(t,x^i) + \delta v^i(t,x^i), \\
\mu(t,x^i) & \longrightarrow \mu(t,x^i) + \delta \mu(t,x^i) \nonumber ,  
\end{align}
where the $\delta T(t,x^i)$ etc. are of first order in derivatives, do not affect the ideal part of the theory, but do introduce shifts at first order. Physically this means that we may consider a family of redefinitions of the hydrodynamic variables $T(t,x^i),v^i(t,x^i),\mu(t,x^i)$ by gradient terms which all agree in equilibrium, when those gradient terms vanish. Such a redefinition, by the chain rule, also causes shifts of the thermodynamic functions $\mathcal{E},P,\rho,n$.

The second kind of ambiguity arises since, when working to first order in the gradient expansion, one may always use the zeroth-order equations of motion in order to simplify the expressions appearing at first order. 
\subsubsection{Tensor structures and equations of motion ambiguity}
\begin{table}[t]
\renewcommand*{\arraystretch}{1.4}
\centering
\begin{tabular}{|c|c| }
  \hline
   scalars &$\left[v^{k}\partial_{k} T\right]$, \,$ v^k \partial_k v^2$, \, $v^k \partial_k \frac{\mu}{T}$,\, $\left[\partial_t T\right]$, \,  $\partial_t v^2$, \, $\left[\partial_t \frac{\mu}{T}\right]$, \, $\partial_k v^k$ \\
  \hline
 vectors & $\left[\partial_i T\right]$, \, $\partial_i v^2$, \, $\partial_i \frac{\mu}{T}$, \, $\partial_t v^i$, \, $v^k \partial_k v^i$, \, $v^i \cdot$ (scalars) \\
  \hline
  tensors & $\sigma_{ij}$, \, $v^{(i} \cdot$ (vectors)$^{j)}$, 
   \, $\delta^i_j \cdot$ (scalars)\\
  \hline
\end{tabular}
\caption{An overview of all the different one-derivative terms of the thermodynamic variables we can write down, with respect to the symmetries of the system. The parentheses around the indices denote a symmetric tensor, since $T^{i}\!_{j} = T^{j}\!_{i}$. The terms in square brackets are the terms we eliminate using the equations of motion at zeroth order. Here $\sigma_{ij}=\partial_i v^j + \partial_j v^i - \delta^i\!_j\frac{2}{d}\partial_k v^k$ is the shear tensor and $d$ is the number of spatial dimensions. Note that we do not decompose the vectors into components transverse to the velocity, since we are also interested in situations where $v^i=0$.}
\label{Tab:tableStructures}
\end{table}
Concretely this is done by using the equations to express certain tensor structures in terms of the remaining ones. The most efficient way to implement this, is by first enumerating all possible tensor structures allowed by the symmetries at first order and then eliminating a convenient set using the zeroth order equations of motion. This then results in a smaller effective set of tensor structures, which are then still subject to the field redefinition ambiguity mentioned above. However, it is considerably simpler to apply the latter to the reduced effective set of tensor structures, which is the procedure we will follow.
Indicated in table \ref{Tab:tableStructures} are all the allowed tensor structures at first order, classified by their content under the ISO$(d)$ symmetry. In this work we focus solely on parity-preserving transport. Note it is natural to consider further decomposing the vectors listed in \ref{Tab:tableStructures} into scalars (by contracting with $v^i$) and pieces transverse to the velocity (by using a projector $P^{ij} = \delta^{ij} - v^iv^j/v^2$). 
However, we opt not to take this step since such a decomposition fails to be well-defined for $v^i=0$. This is to be contrasted with the relativistic case where one may always decompose with respect to $u^\mu$.

Note that the same tensor structure can appear multiple times in different currents. For example, the scalars are counted seven times, because they contribute five times to the stress tensor, and two times to the U$(1)$ current. To be more precise, they appear once in $J^0$ and once in $J^i$, where they are multiplied by $v^i$ to form a vector. Similarly they contribute to different index structures in the stress tensor multiplied by appropriate combinations of the $v^i$ and $\delta^{ij}$. 

Having thus defined the full set of tensor structures, we use the equations of motion at the ideal fluid level,
\begin{align}
\partial_{\mu} T^{(0)}{}^{\mu}\!_{0} & = 0, \\
\partial_{\mu} T^{(0)}{}^{\mu}\!_{j} & = 0, \\
\partial_{\mu} J^{(0)}{}^{\mu}& = 0, 
\end{align}
to eliminate as many first-order structures as possible. We have of course a large amount of freedom to choose which ones to eliminate. We go with the customary selection of eliminating $\partial_iT, \partial_t T$ and $\partial_t\frac{\mu}{T}$. This gives us the desired reduced set of  physical tensor structures, in which to expand our field-redefinitions. 

We now have all the ingredients necessary to write down and constrain the first-order stress tensor and current.
\subsubsection{First-order constitutive relations and choice of frame}
Expanded in our reduced basis of tensor structures (see table \ref{Tab:tableStructures}), the field redefinitions now take the form
\begin{align}
\delta T &\longrightarrow   a_1 v^k \partial_k v^2 + a_2 v^k \partial_k \frac{\mu}{T} + a_3 \partial_t v^2  + a_4 \partial_k v^k  \\
\delta \mu &\longrightarrow a_5 v^k \partial_k v^2 + a_6 v^k \partial_k \frac{\mu}{T} + a_7 \partial_t v^2  + a_8 \partial_k v^k  \\
\delta v^i &\longrightarrow   a_{9} \partial_i v^2 + a_{10} \partial_i \frac{\mu}{T} + a_{11} \partial_t v^i + a_{12} v^k\partial_{k} v^i  + \nonumber \\
&+ v^i \left(a_{13} v^k \partial_k v^2 + a_{14} v^k \partial_k \frac{\mu}{T} + a_{15} \partial_t v^2  + a_{16} \partial_k v^k  \right) 
\end{align}
with free coefficients $\left\{ a_i \right\}_{i=1}^{16}$.  These redefinition ambiguities see themselves confronted with the most general first order stress tensor, again expanded in the reduced basis of tensor structures
\begin{align}
\Pi^{0}\!_{0} &= c_1 v^k \partial_k v^2 + c_2 v^k \partial_k \frac{\mu}{T} + c_3 \partial_t v^2  + c_4 \partial_k v^k \label{eq:T00before}  \\
\Pi^{0}\!_{j} &= c_{5} \partial_j v^2 + c_{6} \partial_j \frac{\mu}{T} + c_{7} \partial_t v^j + c_{8} v^k\partial_{k} v^j  + \nonumber \\
&+ v^j \left(c_{9} v^k \partial_k v^2 + c_{10} v^k \partial_k \frac{\mu}{T} + c_{11} \partial_t v^2  + c_{12} \partial_k v^k  \right) \\
\Pi^{i}\!_{0} &= c_{13} \partial_i v^2 + c_{14} \partial_i \frac{\mu}{T} + c_{15} \partial_t v^i + c_{16} v^k\partial_{k} v^i  + \nonumber \\
&+ v^i \left(c_{17} v^k \partial_k v^2 + c_{18} v^k \partial_k \frac{\mu}{T} + c_{19} \partial_t v^2  + c_{20} \partial_k v^k  \right) \\
\Pi^{i}\!_{j} &= c_{21} \sigma_{ij} + c_{22} \left(v^i\partial_j v^2 + v^j \partial_i v^2 \right) + c_{23} \left(v^i \partial_j \frac{\mu}{T} + v^j \partial_i \frac{\mu}{T} \right) \nonumber \\
 &+  c_{24} \left(v^i \partial_t v^j + v^j \partial_t v^i \right) + c_{25} \left(v^i v^k\partial_k v^j + v^j v^k\partial_k v^i \right) \nonumber \\ 
&+ v^i v^j \left(c_{26} v^k \partial_k v^2 + c_{27} v^k \partial_k \frac{\mu}{T} + c_{28} \partial_t v^2  + c_{29} \partial_k v^k \right)   \nonumber \\
& + \delta^{i}_{j} \left(c_{30} v^k \partial_k v^2 + c_{31} v^k \partial_k \frac{\mu}{T} + c_{32} \partial_t v^2  + c_{33} \partial_k v^k  \right) 
\end{align}

containing the set $\left\{ c_i \right\}_{i=1}^{33}$ of unconstrained coefficients. To this we add the current

\begin{align}
\Pi^{0} &= c_{34} v^k \partial_k v^2 + c_{35} v^k \partial_k \frac{\mu}{T} + c_{36} \partial_t v^2  + c_{37} \partial_k v^k \\
\Pi^{i} &= c_{38} \partial_i v^2 + c_{39} \partial_i \frac{\mu}{T} + c_{40} \partial_t v^i + c_{41} v^k\partial_{k} v^i  + \nonumber \\
&+ v^i \left(c_{42} v^k \partial_k v^2 + c_{43} v^k \partial_k \frac{\mu}{T} + c_{44} \partial_t v^2  + c_{45} \partial_k v^k  \right) \label{eq:Jibefore}
\end{align}
which carries the set $\left\{ c_i \right\}_{i=34}^{45}$ of unconstrained coefficients, which, using the ambiguities above get whittled down to $29$ remaining physical transport coefficients. Imposing that non-dissipative contributions vanish may further reduce this number and we return to this point in our analysis of the entropy current. This is still a somewhat large number, but in the following we will develop more intuition for their physical meaning by considering limiting cases with more symmetries.

Implementing the zeroth-order shifts results in
\bea
T^{\mu}{}_\nu &=& T^{(0)}{}^{\mu}{}_\nu + \underbrace{ \delta \left(T^{(0)}{}^{\mu}{}_\nu \right) + \Pi^{\mu}{}_\nu  \left( \left\{ c_i  \right\} \right)}_{ = \Pi{}^{\mu}{}_\nu  \left( \left\{ \tilde c_i  \right\} \right)}\\
J^{\mu} &=& J^{(0)}{}^{\mu}  + \underbrace{\delta \left(J^{(0)}{}^{\mu} \right) + \Pi^{\mu} \left( \left\{ c_i  \right\} \right)}_{ = \Pi^{\mu} \left( \left\{ \tilde c_i  \right\} \right)}\,,
\eea
in other words we may think of the field redefinitions as acting on the first-order stress tensor and current by moving in the space of coefficients 
\be 
c_i \rightarrow \tilde{c_i}  = c_i + M_{ij} a^j \,,
\ee
where the coefficients $M_{ij}$ form the elements of a $45\times 16$ matrix.

The structure we just exposed means that the ambiguities of the first-order stress tensor and current can be understood as a linear, $\alpha$-dependent trajectory in the space of coefficients $\{ c_i \}$. Any two stress tensors lying on the same trajectory through this space are physically equivalent (and similarly for the current), and so our next goal is to put conditions on the first-order stress tensor and current that fix the ambiguity. In other words, we would like to select one representative for each orbit through the space of $\{c_i\}$. We will now demonstrate that the natural generalization of what is usually called the Landau frame supplies a sufficient number of conditions on the first-order quantities to fully fix their form. The Landau frame conditions appropriate to our symmetry class were given in \cite{deBoer:2017abi} and read
\begin{equation}\label{eq:LandauT} 
T^{\mu}\!_{\nu} u^{\nu} = - \tilde{\mathcal{E}} u^{\mu} \Longrightarrow \begin{cases}
\mu = 0: &\quad T^{0}\,_{0} + T^{0}\,_{j} v^{j} = - \tilde{\mathcal{E}} \\
\mu = i: &\quad T^{i}\,_{0} + T^{i}\,_{j} v^j = - \tilde{\mathcal{E}}v^i 
\end{cases}
\end{equation}
The eigenvector $u^{\mu}$ is parametrized as $u^{\mu} = u^0 (1, v^i)$, where $u^0$ is a function of thermodynamic variables and is not fixed. The frame condition for the $U(1)$ current can be taken as \cite{deBoer:2017ing}
\begin{equation}\label{eq:LandauJ}
\bar{u}_{\mu} J^{\mu} =-\frac{1}{u^0} n. 
\end{equation}
In fact $u^0$ is not the only freedom that enters in the choice of frame.  Due to the lack of a metric structure to lower the index on $u^\mu$, the corresponding object with index down, $\bar{u}_{\mu}$, is again for us to choose. We shall employ the same choice as \cite{deBoer:2017ing}, namely we take $\bar{u}_{\mu}$, such that
\be
u^{\mu}\bar{u}_{\mu} = -c^2\,,\qquad \textrm{with}\qquad \bar{u}_{\mu} = \frac{1}{u^0}(-c^2-B v^2, B v^i)
\ee
parametrized in terms of $v^i$ and the free function $B$, which may in general depend on the thermodynamic variables.  The parameters $B, c^2$ and $u^0$ are all free choices giving different hydrodynamic frames, that is for each  choice of $B, c^2, u^0$ there is a corresponding Landau frame. However, we retain them explicitly in our construction, as this facilitates our later analysis of limiting cases.  In certain more highly symmetric situations these parameters are naturally chosen by the requirement that the frame condition is constructed in accordance with the symmetries at hand. For example for a Lorentz invariant fluid $(u^{0})^2=B=\frac{1}{1-v^2/c^2}$, while a Galilean invariant fluid has $B=0$ and $u^0=1$.

Fixing the frame will reduce the number of coefficients we have in the constitutive relations. Initially, we had 45 different coefficients \{$c_i$\} in \eqref{eq:T00before} - \eqref{eq:Jibefore}. The Landau frame conditions, \eqref{eq:LandauT} and \eqref{eq:LandauJ}, give us 16 constraints, resulting in 29 physical transport coefficients. In practice, we solve the Landau conditions by requiring all the coefficients that multiply different tensor structures in different components of $\Pi^{\mu}\!_{\nu}$ and $J^{\mu}$ to vanish. These coefficients are functions of $c_i$ and $a_i$, and we solve the resulting 16 equations for $a_1,\dots, a_{16}$. Having done that, the new coefficients are functions of only $c_i$ and we can explicitly see that there are only 29 different coefficients. These we have relabelled as detailed below, and each are functions of $T, v^2, \mu$. This brings us to our main result for the constitutive relations at first order.

\be
\def\arraystretch{1.2}
\boxed{
\begin{array}{lll}
\textbf{Stress tensor}\span\span\\
\Pi^{0}\!_{0} &=& \gamma_2 v^k\partial_kv^2  
+ \left(\gamma_1 v^2 + \frac{\bar{\pi}}{2}\right) \partial_t v^2  
+ \gamma_3 v^2 \partial_k v^k  
+ (\gamma_4 v^2 - T(\bar{\alpha}+\bar{\gamma})) v^k\partial_k \frac{\mu}{T}  \vspace{1em}\\
\Pi^{0}\!_{j} &=& \gamma_{5} \partial_j v^2 
+ \gamma_{6} v^k\partial_k v^j 
+ T (\bar{\alpha}+ \bar{\gamma}) \partial_j \frac{\mu}{T} 
-\bar{\pi} \partial_t v^j \\
&&-v^j\left(
\frac{1}{2 v^2} (2\gamma_2 + 2\gamma_5 + \gamma_6)v^k\partial_kv^2
-\gamma_4 v^k \partial_k \frac{\mu}{T} 
-\gamma_1\partial_t v^2
-\gamma_3 \partial_k v^k\right)  \vspace{1em}\\
\Pi^{i}\!_{0}\ &=& \left(\gamma_{8}v^2 + \frac{\bar{\eta}}{2}\right) \partial_i v^2
+\left(\gamma_{13}v^2+\bar{\eta}\right) v^k\partial_{k} v^i
+\gamma_{7} v^2 \partial_t v^i 
+ \gamma_{14} v^2 \partial_i \frac{\mu}{T}\\
&&+ v^i \left(
\gamma_{9} v^k \partial_k v^2 
+ \gamma_{10} \partial_t v^2  
+ \left(\gamma_{11}v^2 + \bar{\zeta} - \frac{2 \bar{\eta}}{d}\right) \partial_k v^k   
+ \gamma_{12} v^k \partial_k \frac{\mu}{T} 
\right)\vspace{1em}\\
\Pi^{i}\!_{j} &=&  \delta^{i}_{j} \left(
\gamma_{15} v^k \partial_k v^2 
+ \gamma_{17} v^k \partial_k \frac{\mu}{T} 
+ \gamma_{16} \partial_t v^2  
-\bar{\zeta} \partial_k v^k  
\right) 
-\bar{\eta} \sigma_{ij} \\
&&-\gamma_8 \left(v^i\partial_j v^2 + v^j \partial_i v^2 \right) 
-\gamma_{14} \left(v^i \partial_j \frac{\mu}{T} + v^j \partial_i \frac{\mu}{T} \right) \\
&& -\gamma_{7} \left(v^i \partial_t v^j + v^j \partial_t v^i \right) 
-\gamma_{13} \left(v^i v^k\partial_k v^j + v^j v^k\partial_k v^i \right) \\ 
&&+ \frac{v^i v^j}{v^2} \left( 
(\frac{1}{2}\gamma_{13} - \gamma_{15} + \gamma_8 -  \gamma_9) v^k \partial_k v^2 
-(\gamma_{12}-\gamma_{14}+\gamma_{17}) v^k \partial_k \frac{\mu}{T} \right.\\
 &&\left.-(\gamma_{10} + \gamma_{16}-\frac{1}{2}\gamma_7) \partial_t v^2 
 -\gamma_{11} v^2 \partial_k v^k \right) \vspace{1em}\\
\textbf{U(1) current}\span\span\\
\Pi^0 &=& \frac{B}{c^2+B v^2} \left(
\gamma_{19} v^k \partial_k v^2 
+\frac{1}{2}(\bar{\alpha} + 2 v^2 \gamma_{18} - \bar{\gamma})\partial_t v^2  
+ \gamma_{20} v^2 \partial_k v^k 
+ (\gamma_{21} v^2 - T\bar{\sigma}) v^k \partial_k \frac{\mu}{T} 
\right)\vspace{1em} \\
\Pi^{i}  &=& \gamma_{22} \partial_i v^2 
+ \gamma_{23} v^k\partial_{k} v^i  
- T \bar{\sigma} \partial_i \frac{\mu}{T} 
+ (\bar{\alpha} - \bar{\gamma}) \partial_t v^i \\
&&+ v^i \left(  \frac{1}{2 v^2} (2\gamma_{19} - 2 \gamma_{22} - \gamma_{23}) v^k \partial_k v^2 
+ \gamma_{21} v^k \partial_k \frac{\mu}{T} +\gamma_{18} \partial_t v^2 + \gamma_{20} \partial_k v^k  \right) \vspace{1em}\\
\textbf{Transport coefficients}\span\span\\
\bar{\eta}, \bar{\zeta}, \bar{\sigma}, \bar{\alpha}, \bar{\gamma}, \bar{\pi}, \quad \gamma_1,\ldots,\gamma_{23}\qquad \text{(each a function of $T,v^2,\mu$)}\span\span
\end{array}
}\label{constitutive}
\ee
Note that 29 transport coefficients is the count before the constraints of positivity of entropy current have been applied.
The transport coefficients $\bar{\eta}, \bar{\zeta}, \bar{\sigma}, \bar{\alpha}, \bar{\gamma}, \bar{\pi}$ become those utilised in \cite{deBoer:2017abi} in a linear perturbation analysis around a uniform zero-velocity background, i.e.  $\eta_0, \zeta_0, \sigma_0, \alpha_0, \gamma_0, \pi_0$, respectively, but otherwise differ as here they are functions of $v^2$ too. We will see that imposing boost symmetry, be it Lorentz, or Galilean,  significantly reduces the number of transport coefficients. Imposing Lifshitz symmetry will constrain the functional form of the transport coefficients as well as reduce their number (though not as significantly as boost symmetry).

\subsection{Entropy current}
One of the physical requirements of any theory of hydrodynamics is its adherence to the second law, namely the positivity of entropy production. At equilibrium one can readily identify a unique entropy current, however at first order in the hydrodynamic expansion there can be several ambiguities. Nevertheless, as is well-known, the process of defining the most general entropy current and demanding that its divergence is non-negative can still lead to definitive constraints on transport coefficients. We shall elucidate this in what follows.

The procedure outlined above is straightforward; we wish to construct the most general expression for the entropy current $S^\mu$ consistent with the symmetries at hand, up to first derivative order, and then ensure that $\partial_\mu S^\mu \geq 0$. 
The entropy current at ideal order is then given by $S^\mu = s v^\mu + O(\partial)$ where $v^\mu \equiv (1,v^i)^\mu$.
One can verify that with this definition, $\partial_\mu S^\mu  = 0 + O(\partial)^2$, ensuring there is no entropy production at this hydrodynamic order.

In order to construct the most general first order contribution to $S^\mu$ consistent with symmetries, it is convenient to split $S^\mu$ into two contributions, a canonical part, and a non-canonical part,
\be
S^\mu = S_{\rm can}^\mu + S_{\rm non}^\mu.
\ee
We shall begin with the canonical part. Consider the expression for the entropy density,
\bea
T s &=&  \varepsilon - \rho v^2 + P - \mu n \\
\implies T s v^\mu &=& -T^{(0)}{}^\mu{}_\nu v^\nu + P v^\mu  - \mu J^{(0)}{}^\mu.\label{sthermo}
\eea
Inspired by \eqref{sthermo} we now define $S_{\rm can}^\mu$ which, by construction, differs from $s v^\mu$ by terms that are at least first order in derivatives, 
\be
T S_{\rm can}^\mu \equiv -T^\mu{}_\nu v^\nu + P v^\mu  - \mu J^\mu,
\ee
whose divergence is easily evaluated, 
\be
\partial_\mu S_{\rm can}^\mu = - \Pi^\mu{}_\nu\partial_\mu \frac{v^\nu}{T} - \Pi^\mu \partial_\mu \frac{\mu}{T} \label{divScan}
\ee
and crucially depends only on products of first derivatives; there are no explicit second derivative terms. What remains in order to construct $S^\mu$ proper is the non-canonical part, which is simply the most general set of terms consistent with the symmetries at hand. In other words, the construction of $S_{\rm non}^\mu$ parallels our enumeration of the possible terms allowed in the constitutive relations:
\bea
S_{\rm non}^{0} &=& \tilde{c}_{1} v^k \partial_k v^2 + \tilde{c}_{2} v^k \partial_k \frac{\mu}{T} + \tilde{c}_{3} \partial_t v^2  + \tilde{c}_{4} \partial_k v^k, \label{Snon0}\\
S_{\rm non}^{i}  &=& \tilde{c}_{5} \partial_i v^2 + \tilde{c}_{6} \partial_i \frac{\mu}{T} + \tilde{c}_{7} \partial_t v^i + \tilde{c}_{8} v^k\partial_{k} v^i  + \nonumber \\
&&+ v^i \left(\tilde{c}_{9} v^k \partial_k v^2 + \tilde{c}_{10} v^k \partial_k \frac{\mu}{T} + \tilde{c}_{11} \partial_t v^2  + \tilde{c}_{12} \partial_k v^k  \right).\label{Snoni}
\eea
There are two types of contribution to $\partial_\mu S^\mu$: there are products of first derivatives (as is the case for $\partial_\mu S_{\rm can}^\mu$), and there are second derivative terms. Our first task is to use the equations of motion to maximally reduce the number of terms that can appear, and we do so here by eliminating all terms that contain one or more time derivatives. We then require that the coefficients of the remaining second derivative terms vanish, since they do not have a definite sign. This results in equality-type constraints (as opposed to inequality-type) that fix 8 out of the 12 coefficients in the definition of $S^{\mu}_{non}$,
\be
\begin{array}{l | l | l | l}
\tilde{c}_5 = 0 &
\tilde{c}_6 = 0 &
\tilde{c}_{10} = \tilde{c}_2 &
\tilde{c}_3 = m_1 \tilde{c}_2\\
\tilde{c}_4+\tilde{c}_7 =m_2 \tilde{c}_2&
\tilde{c}_8 +\tilde{c}_{12}= m_3 \tilde{c}_2&
\tilde{c}_9 = m_4 \tilde{c}_2&
\tilde{c}_1+\tilde{c}_{11}=  (m_1+m_2)\tilde{c}_2
\end{array}\label{2derivconstraints}
\ee
where $m_{1,2,3,4}$ are constants of proportionality determined completely by the equation of state.\footnote{These coefficients are provided in a companion notebook to the paper.} After imposing these conditions $\partial_\mu S^\mu$ is then quadratic in derivatives, taking the general form,
\be
\partial_\mu S^\mu =
\left(\begin{matrix}
    \partial_i T \\
    \partial_i \mu \\
    \partial_{i} v_{j}
\end{matrix}\right)^{\text{T}}
\left(
\begin{matrix}
    T_{ik} & A_{ik} & B_{ikl} \\
    A_{ki} & M_{ik} & C_{ikl} \\
    B_{kij} & C_{kij} & V_{ijkl}
\end{matrix}
\right)\left(\begin{matrix}
    \partial_k T \\
    \partial_k \mu \\
    \partial_{k} v_{l}
\end{matrix}\right) \label{quadraticform}
\ee
where the components of the matrix are given by all possible index contractions,
\bea
T_{ik} &=& b_0 \delta_{ik} + b_1 v_i v_k\\
A_{ik} &=& b_2 \delta_{ik} + b_3 v_i v_k\\
M_{ik} &=& b_4 \delta_{ik} + b_5 v_i v_k\\
B_{ikl} &=& b_6 v_i \delta_{kl} + b_7 v_i v_k v_l + b_8 \delta_{ik} v_l + b_9\delta_{il} v_k\\
C_{ikl} &=& b_{10} v_i \delta_{kl} + b_{11} v_i v_k v_l + b_{12} \delta_{ik} v_l + b_{13}\delta_{il} v_k\\
V_{ijkl} &=& b_{14} \delta_{ik}\delta_{jl} + b_{15} \delta_{il}\delta_{jk} +b_{16} \delta_{ij} \delta_{kl} + b_{17} v_i v_j v_k v_l\nonumber \\ 
&& b_{18}(v_iv_j\delta_{kl} +  v_k v_l \delta_{ij}) + b_{19} (v_i v_l \delta_{jk} + v_j v_k \delta_{il})+\nonumber \\ 
&& b_{20} v_i v_k\delta_{jl}+ b_{21} v_j v_l \delta_{ik}.
\eea
We have computed each of these $b_I$ coefficients explicitly, and in general they depend on the transport coefficients, the remaining redundancies in the definition of $S_{\rm non}^\mu$ (i.e. the $\tilde{c}$) plus equation of state data.\footnote{These coefficients are provided in a companion notebook to the paper.}   For example, some of the more compact expressions we have encountered are,
\bea
b_{14} &=& \frac{\bar{\eta}}{T} \label{blisting1}\\
b_{15} &=& \tilde{c}_8 + \frac{\bar{\eta}}{T}\\
b_{20} &=& \frac{\bar{\pi} + \gamma_6-\gamma_7+\gamma_{13}}{T} \label{blisting2}\\
\frac{b_{21}}{4}-\frac{b_{19}}{2}+\frac{b_{20}}{4} &=& -\frac{1}{2}\frac{\partial \tilde{c}_3}{\partial v^2}.
\eea
So far, we have ensured that second derivative terms vanish, and computed the resulting quadratic form explicitly \eqref{quadraticform}. We note that there are 22 coefficients appearing in the quadratic form but there are a total of 29 transport coefficients, and so we expect there to be non-dissipative combinations of transport coefficients which we will enumerate in the next section \ref{sec:nd}. 

We also expect an additional class of equality-type constraints on transport coefficients. These usually arise in considerations of couplings to background fields, and by imposing time-reversal covariance in the form of Onsager reciprocal relations. In some cases these may be related to the non-dissipative contributions we discuss below.

Finally, the physical requirement of the second law $\partial_\mu S^\mu \geq 0$ requires that the quadratic form \eqref{quadraticform} be non-negative for all fluid configurations. These typically result in inequality-type constraints. As many of the coefficients appearing in \eqref{quadraticform} are of considerable length, we will not present an exhaustive analysis of these inequalities. However in section \ref{sec:entshear} we present a complete analysis of the inequalities that can be extracted by studying shear-type perturbations of uniform flow.

\subsubsection{Non-dissipative transport coefficients\label{sec:nd}}
There are linearly independent combinations of transport coefficients that do not enter $\partial_\mu S^\mu$. Such terms are therefore responsible for effects which are nonuniform and non-dissipative, and thus potentially interesting physical effects in their own right. Moreover a theory constructed from such terms alone contains an additional conserved current, $S^\mu$, and may therefore not be subject to the usual difficulties in constructing Lagrangian descriptions of hydrodynamics (see, for example, \cite{Bhattacharya:2012zx}). On the other hand, there may be physical requirements which dictate that such terms vanish. For example, constraints on such terms arise in the study of hydrostatic partition functions \cite{Banerjee:2012iz,Jensen:2012jh}. To explore these terms in detail, in this section we enumerate the constraints that must be placed on the general theory to remove all non-dissipative contributions. 

We begin by decomposing the currents appearing in $\partial_\mu S_{\rm can}^\mu$, \eqref{divScan}, into dissipative and non-dissipative pieces,
\bea
\Pi^\mu{}_\nu &=& (\Pi_{\rm D})^\mu{}_\nu+(\Pi_{\rm ND})^\mu{}_\nu,\\
\Pi^\mu &=& (\Pi_{\rm D})^\mu+(\Pi_{\rm ND})^\mu,
\eea
where the non-dissipative pieces (ND) do not enter $\partial_\mu S^\mu$ by definition, and the dissipative pieces (D) are here assumed to take the form,
\be
\partial_\mu S^\mu = - (\Pi_{\rm D})^\mu{}_\nu\partial_\mu \frac{v^\nu}{T} - (\Pi_{\rm D})^\mu \partial_\mu \frac{\mu}{T}. \label{divSnd}
\ee
There are two classes of such non-dissipative transport coefficients.

The first class are those which arise directly in $\partial_\mu S_{\rm non}^\mu$, as these manifestly do not enter \eqref{divSnd}. These, as we have already seen, are constructed from the remaining coefficients in the constitutive relation for $S_{\rm non}^\mu$ after second-derivative constraints are imposed, namely \eqref{2derivconstraints}, and are given by $\tilde{c}_1, \tilde{c}_2, \tilde{c}_4, \tilde{c}_8$. There can be additional equality-type constraints that result from considering new second-derivative terms that arise when background fields are turned on, as was demonstrated in the Lorentz invariant case \cite{Bhattacharya:2011tra}. Having recognised these terms as non-dissipative, for this section we now set $\tilde{c}_1 = \tilde{c}_2 = \tilde{c}_4 = \tilde{c}_8 = 0$ and the remaining entropy current is purely of canonical form. 

We now turn to the second class of non-dissipative transport coefficients.
Having imposed the vanishing of the first class of coefficients, there are 29 transport coefficients remaining (those appearing in the constitutive relations for $\Pi^\mu{}_\nu$ and $\Pi^\mu$) but only 22 independent terms in the quadratic form (the $b_i$) \eqref{quadraticform}. Thus there are additional independent linear combinations of transport coefficients that are non-dissipative. We can now count how many such non-dissipative transport coefficients there are. Denoting a general transport coefficient as $t_I$, we can define a linear map $\cal M$ from the space of transport coefficients to the space of quadratic-form coefficients,
\be
b_i = {\cal M}_{iJ} t_J,
\ee
where $\cal M$ is a $22\times 29$ rectangular matrix. We are interested in how many linearly independent vectors there are in the vector space of transport coefficients that do not contribute to $b_i$. By direct computation we find that the rank of $\cal M$ is 20 and the dimension of its null space is 9. Thus we conclude that there is a vector space of dimension 20 spanned by dissipative transport coefficients (the image of $\cal M$), and a vector space of dimension 9 spanned by non-dissipative transport coefficients (the null space of $\cal M$). The linearly independent non-dissipative coefficients are as follows;
\bea
&& \gamma_{23} + \frac{\gamma_{22}}{2} + T \gamma_{14} - T\gamma_{12},\label{nondissB}\\
&& \gamma_{18} + 2 T \gamma_4 + 2 v^2 \bar{\gamma},\\
&& \gamma_{16} - \gamma_{2} + 2 \gamma_{3},\\
&& \gamma_{15} + 2\gamma_{11} - \gamma_{9},\\
&& \gamma_{10}-\gamma_{2},\\
&& \gamma_{7}+\gamma_{6} + \frac{\gamma_{5}}{2} - \frac{\gamma_{2}}{2},\\
&&\gamma_{20} - \frac{c^2T}{c^2+Bv^2} (\gamma_{17}-\gamma_{12}) + 2 Q_2 v^2(\gamma_{11}+\gamma_{3}) + Q_1 v^2 \bar{\zeta},\\
&&\gamma_{19} + \frac{c^2T}{c^2+Bv^2} 2\gamma_{12} + 2 Q_1 \gamma_{11} + 2 Q_2 v^2(\gamma_{9} + \gamma_{2}),\\
&&\frac{B T}{c^2+Bv^2} \gamma_{4} - Q_1 \gamma_3 - Q_2(\gamma_2 v^2 + \gamma_1) + \bar{\gamma},
\label{nondissE}
\eea
where we have defined the following thermodynamic quantities,
\bea
Q_1 &\equiv & \frac{B}{c^2 + Bv^2}\frac{P_\mu(T P_{TT}+\mu P_{T\mu})-P_T(T P_{T\mu} +\mu P_{\mu\mu})}{T(P_{T\mu}^2 - P_{TT} P_{\mu\mu})},\\
Q_2 &\equiv & \frac{B}{c^2 + Bv^2}\frac{P_{v^2\mu}(T P_{TT}+\mu P_{T\mu})-P_{Tv^2}(T P_{T\mu} +\mu P_{\mu\mu})}{T(P_{T\mu}^2 - P_{TT} P_{\mu\mu})},
\eea
where $P_{\mu} \equiv \partial_{\mu}P,\, P_{v^2\mu} \equiv \partial_{v^2}\partial_\mu P$ and similarly for other derivatives. The orthogonal complement of this space of transport coefficients is purely dissipative, and an analysis of a theory where these are set to zero would be interesting to study further. One simple example is to set $\gamma_{10} = -\gamma_2$ and then set all other transport coefficients to zero. This theory has $\Pi^\mu=0$ while, 
\be
\Pi^0{}_0 = \gamma_2 v^k \partial_k v^2, \quad \Pi^i{}_0 = -\gamma_{2} v^i \partial_t v^2, \quad \Pi^0{}_j = -\gamma_{2}\frac{v^j v^k \partial_k v^2}{v^2},\quad  \Pi^i{}_j = \gamma_{2}\frac{v^i v^j \partial_t v^2}{v^2}
\ee
and is non-dissipative as is easily verified by evaluating the quadratic form.

As a cautionary remark, the above procedure has to be repeated in cases where additional linear constraints are imposed, such as those arising due to enhanced symmetry. The calculation of the null space should be performed only after additional constraints have been imposed. The reason is that non-dissipative terms in the general theory may not respect those symmetries and consequently contribute to dissipative transport in the more symmetrical theory. On the other hand, once the dissipative terms have been computed as the orthogonal complement of \eqref{nondissB}-\eqref{nondissE}, then one may assess whether or not dissipative terms in a more symmetrical theory are zero, simply evaluating them on the particular transport coefficients of the theory in question. This is the same as computing the quadratic form coefficients.

Before concluding, we can check our non-dissipative constraints for perturbations around $v^i=0$ flows. In this case, the dimension of the null space is much higher, since there are many fewer terms that can appear in the quadratic form. In particular, the null space is enlarged to include all $\gamma_I$ as independent basis vectors. Hence setting all these to zero, leaves a single constraint in \eqref{nondissB}-\eqref{nondissE}, namely $\bar{\gamma}(T,v^2=0,\mu) = \gamma_0(T,\mu) = 0$. This constraint was also found in \cite{deBoer:2017abi} but based on imposing Onsager reciprocity. The connection between Onsager reciprocity and non-dissipative coefficients is manifest when the antisymmetric components of the conductivity matrix are linearly independent of all other contributions. However the interaction between strictly dissipative coefficients, strictly non-dissipative coefficients and those coefficients which are required to vanish due to Onsager reciprocity is not clear in this most general case. In order to understand the role of Onsager and the requirements of microscopic time reversal invariance we would need to conduct an analysis of modes and associated Green's functions along the lines of \cite{Kovtun:2012rj}, which we postpone to future work.

In summary, if we require that all non-dissipative contributions vanish (these are given in \eqref{nondissB}-\eqref{nondissE}), the number of transport coefficients appearing in the stress tensor and U(1) current is reduced from 29 to 20. 

\subsubsection{Example: shear modes and shear viscosity \label{sec:entshear}}
Consider the fluid configuration corresponding to a shear-type velocity perturbation around uniform flow,
\be
T(t,\vec{x})= \bar{T}, \qquad \mu(t,\vec{x}) = \bar{\mu},\qquad v^i = \bar{v}^i  +  \delta v^i(t,k\cdot\vec{x}) \label{shearpert}
\ee
where $k^i$ is a spatial wavevector and $k\cdot \delta v = \bar{v}\cdot \delta v = 0$. For this perturbation the only contributions to the quadratic form \eqref{quadraticform} at second order in the perturbation are,
\bea
\partial_\mu S^\mu &=& b_{14}(T,v^2,\mu) \partial_i v^j \partial_i v^j + b_{20}(T,v^2,\mu) v^j v^k \partial_j v^i \partial_k v^i\\
&=& (b_{14}(\bar{T},\bar{v}^2,\bar{\mu})k^2 + b_{20}(\bar{T},\bar{v}^2,\bar{\mu})(k\cdot\bar{v})^2 )(\partial_2\delta v^i)^2\\
&=& \left(b_{14}(\bar{T},\bar{v}^2,\bar{\mu}) + \bar{v}^2b_{20}(\bar{T},\bar{v}^2,\bar{\mu})(\cos\theta)^2 \right)k^2(\partial_2\delta v^i)^2.
\eea
where $\theta$ is the angle between $k$ and $\bar{v}$ and $\partial_2$ denotes derivative with respect to the second argument of the function $\delta v^i$. Hence for perturbations satisfying $\theta=\pi/2$, positivity of entropy production requires
\be
b_{14}(T,v^2,\mu) \geq 0.
\ee
Positive $b_{14}$ allows $b_{20}$ to take on negative values, with $\partial_\mu S^\mu$ minimised at $\theta = 0$, hence we also require
\be
b_{14}(T,v^2,\mu) +v^2 b_{20}(T,v^2,\mu)\geq 0.
\ee
In terms of transport coefficients listed above, \eqref{blisting1} and \eqref{blisting2}, these constraints become
\be
\boxed{
\begin{array}{ll}
&\bar{\eta} \geq 0,\\
& \bar{\eta} + v^2 (\bar{\pi} + \gamma_6-\gamma_7+\gamma_{13}) \geq 0.\label{shearentropy}
\end{array}
}
\ee
We will show that these constraints coincide with those arising out of the requirement of dynamical stability. Finally, we shall see later in section \ref{sec:lorentz} that for theories with Lorentz boost invariance the two constraints \eqref{shearentropy} coincide, becoming $\eta \geq 0$ where $\eta$ is the usual shear viscosity transport coefficient of relativistic hydrodynamics. Additionally it coincides with the conclusion of $\eta_0\geq 0$ reached in \cite{deBoer:2017abi} where $\eta_0 = \lim_{v^2\to0} \bar{\eta}$.

\subsection{Hydrodynamic shear mode\label{sec:shear}}
In this section we consider hydrodynamic modes, that is, perturbations of a background uniform flow that satisfy the hydrodynamic conservation equations. They describe how small departures from uniformity relax over time (quasinormal modes), or how they decay spatially in the context of non-equilibrium steady states (spatial collective modes). We consider a background temperature $\bar{T}$ chemical potential $\bar{\mu}$ and velocity $\bar{v}^i$, and consider the equations of motion resulting from the general first order constitutive relation \eqref{constitutive}. Specifically we focus on the shear-type perturbation \eqref{shearpert} in Fourier space, $\delta v^i(t,k\cdot \vec{x}) = e^{-i \omega t + i k\cdot \vec{x}} \delta v^i$. 
Such perturbations are also hydrodynamic modes provided a dispersion relation $\omega(k^i)$ is satisfied. To linear order in amplitude, and to first hydrodynamic order, the constitutive relations are perturbed as follows,
\bea
\delta T^0{}_j &=& e^{-i \omega t + i k\cdot \vec{x}}  \left(\rho + ik\cdot\bar{v} \gamma_6 + i\omega\bar{\pi}  \right) \delta v^j, \\ 
\delta T^i{}_j &=& e^{-i \omega t + i k\cdot \vec{x}} \left((\rho - i k\cdot\bar{v} \gamma_{13} + i \omega \gamma_7) (\bar{v}^i \delta v^j + \bar{v}^j \delta v^i) - i \bar{\eta} (k^i \delta v^j + k^j \delta v^i)\right)
\eea
and the equations of motion thus give rise to the following dispersion relation, as a Taylor series in gradients,
\be
i \rho(k\cdot\bar{v} - \omega) + (\bar{\pi}\omega^2 + (\gamma_6-\gamma_7) k\cdot\bar{v} \omega + \bar{\eta}k^2 + (k\cdot\bar{v})^2\gamma_{13})  + O(\omega,k)^3 = 0.
\ee
There are two roots of this polynomial, $\omega(k)$, however one root $\omega = i \rho/\bar{\pi}+ O(k)$ is not consistent with the hydrodynamic gradient expansion and we discard it. The other root is,
\be
\omega(k^i) = k\cdot\bar{v}  - i \frac{\bar{\eta} k^2 + (\bar{\pi} + \gamma_6 - \gamma_7 + \gamma_{13}) (k\cdot\bar{v})^2}{\rho} + O(k)^3. \label{sheardispersion}
\ee
We observe that this hydrodynamic mode is stable, with a frequency in the lower-half complex plane, provided the conditions derived from positivity of entropy production are met, \eqref{shearentropy}. Therefore the conditions of dynamical stability and positivity of entropy production coincide here. We also note that all combinations of transport coefficients entering here are in the orthogonal complement of the purely non-dissipative sector \eqref{nondissB}-\eqref{nondissE} and so in a theory with only non-dissipative terms these $O(k)^2$ pieces vanish.

So far we considered the requirements according to $\omega \in \mathbb{C}$ with $k\in \mathbb{R}^d$. Of recent physical interest are modes with complex momenta; spatial collective modes which can obtained from this dispersion relation by fixing $\omega =0$ (or more generally, $\omega \in \mathbb{R}$) and continuing to complex momenta $k\in \mathbb{C}^d$, yielding a dispersion relation of the form $k^i(\bar{v}^j)$ describing how decay lengths in stationary systems depend on background velocity \cite{Sonner:2017jcf,Novak:2018pnv}.

From \eqref{sheardispersion} we can see that the mode is purely diffusive if we move to a coordinate system that comoves with the fluid at velocity $\bar{v}$, but as the explicit $\bar{v}^2$ dependence makes clear, the different $\bar{v}$ frames are physically inequivalent as expected due to the lack of boost invariance. We can see a particular combination of the 29 transport coefficients enter this dispersion relation; it would be interesting to analyse the physical consequences of the other new transport coefficients through studying hydrodynamic modes in other sectors: sound and charge diffusion.

\section{Special cases}

\subsection{Lorentz boosts\label{sec:lorentz}}
Theories admitting Lorentz invariance are a special case of our general $\mathbb{R}_t\times$ISO$(d)\times$U$(1)$ construction,  enlarging the number of symmetries. Imposing such additional symmetry requirements on our constitutive relations  \eqref{constitutive} severely constrains the allowed form of the 29 transport coefficients. 
In this section we will calculate these 29 coefficients for a Lorentz invariant theory, and show how they are completely determined by only 4 free functions of two variables. This is further reduced to 3 after imposing positivity entropy production. These are of course the well-known transport coefficients of first order relativistic hydrodynamics.

A Lorentz boost by a velocity $c\beta^i$, with respect to the speed of light $c$, is achieved by the following coordinate transformation,
\be
t\to t + \frac{\beta_i x^i}{c}, \quad x^i \to x^i + \beta^i c t, \label{LorentzBoostCoords}
\ee
working to linear order in $\beta^i$, the velocity transforms as
\be
v^i \to v^i +  c \beta^i - \frac{\beta\cdot v}{c} v^i. \label{LorentzBoostVel}
\ee
In addition, as we shall see, we also require that the following quantities are invariant in order that we have a non-trivial equation of state,
\be
\tilde{T} \equiv \gamma T, \qquad \tilde{\mu} \equiv \gamma\mu \label{TtildeDef}
\ee
where we have introduced the Lorentz factor $\gamma \equiv \left(1-v^2/c^2\right)^{-1/2}$, and which completes the transformation rules for all hydrodynamic variables, 
\be
T \to T - \frac{\beta\cdot v}{c} T,\qquad  \mu \to \mu - \frac{\beta\cdot v}{c} \mu. \label{LorentzBoostTM}
\ee
We require that the stress tensor and U(1) current transform as Lorentz tensors under the above linearised transformations, for any boost parameter $\beta^i$. This gives rise to a set of equations that must be satisfied, leading to constraints on both the thermodynamic variables and the transport coefficients. In more detail, recall that all thermodynamic variables and transport coefficients are functions of $T,v^2,\mu$, and so performing the above boost \eqref{LorentzBoostCoords} -- which affects the hydrodynamic variables through \eqref{LorentzBoostVel} and \eqref{LorentzBoostTM} -- leads to a Taylor expansion of the transport coefficients to order $\beta^i$. Thus, demanding the correct transformation rule under any linear boost parameter $\beta^i$ results in a set of partial differential equations for the transport coefficients in terms of $T,v^2,\mu$, the solutions to which we will set out in the two subsections that now follow.

\subsubsection{Ideal hydrodynamic order}
At ideal hydrodynamic order, by demanding that the component $T^{(0)}{}^0{}_j$ transform correctly, we find the following constraint (as a coefficient of the parameter of the boost $\beta^j$),
\be
\rho c^2 = P+\mathcal{E}. 
\ee
This is so far independent of any transformation rules for $T,\mu$, which cannot contribute to the term proportional to $\beta^j$.
Upon utilising thermodynamic identities this gives the following PDE for the equation of state,
\be
\left(\mu \partial_\mu  - 2 (c^2-v^2)\partial_{v^2} + T\partial_T\right) P(T,v^2,\mu) = 0
\ee
with a general solution
\be
P = \tilde{P}\left(\gamma T, \gamma \mu\right) = \tilde{P}\left(\tilde{T},\tilde{\mu}\right). \label{Ptilde}
\ee
where $\tilde{T},\tilde{\mu}$ are defined in \eqref{TtildeDef}.
All remaining PDEs resulting from demanding Lorentz invariance at ideal order are now solved by \eqref{Ptilde}, provided we also ensure that $T,\mu$ transform leaving $\tilde{T},\tilde{\mu}$ invariant, i.e. under the rule \eqref{LorentzBoostTM}. If we do not, then additional constraints arise on $\tilde{P}$ and prevent us from having a general function of two independent variables. Furthermore demanding that the frame conditions \eqref{eq:LandauT} and \eqref{eq:LandauJ} take Lorentz covariant form determines
\be
B = \gamma^2, \qquad u^0 = \gamma \label{lorframe}
\ee
where we have also fixed an arbitrary constant of proportionality in $u^0$. \eqref{lorframe} furnishes us with a covariant vector $u^\mu$ which we can now use to construct covariant forms of the constitutive relations in the usual way.
With \eqref{Ptilde} imposed, we arrive at the following familiar constitutive relations at ideal order,
\bea
T^{(0)}{}^\mu{}_\nu &=& \tilde{\cal E}\frac{u^\mu u_\nu}{c^2} + \tilde{P} \Delta^\mu_{~\nu}\\
J^{(0)}{}^\mu &=& \tilde{n} u^\mu
\eea
where we have introduced the projector,
\be
\Delta^\mu_{~\nu} = \delta^\mu_\nu + \frac{1}{c^2} u^\mu u_\nu
\ee
and the thermodynamic relations,
\bea
\tilde{s} &=& \partial_{\tilde{T}}\tilde{P},\\
\tilde{n} &=& \partial_{\tilde{\mu}}\tilde{P},\\
\tilde{\cal E} &=& -\tilde{P} + \tilde{s} \tilde{T} + \tilde{n} \tilde{\mu}.
\eea

\subsubsection{First order}
At first order, after solving the multitudinous PDEs, we arrive at the following expressions for the first $6$ transport coefficients
\bea
\bar{\eta}(T,v^2,\mu) &=& \gamma\,\eta(\tilde{T},\tilde{\mu})\\
\bar{\zeta}(T,v^2,\mu) &=& \gamma\,\zeta(\tilde{T}, \tilde{\mu})\\
\bar{\pi}(T,v^2,\mu) &=& \frac{\gamma^3 v^2}{c^4}\,\eta(\tilde{T},\tilde{\mu})\\
\bar{\alpha}(T,v^2,\mu) &=& -\frac{\gamma^2}{2c^2}\tilde{T}\,\chi(\tilde{T},\tilde{\mu})\\
\bar{\gamma}(T,v^2,\mu) &=& \frac{\gamma^2}{2c^2}\tilde{T}\,\chi(\tilde{T},\tilde{\mu})\label{barGammaLor}\\
\bar{\sigma}(T,v^2,\mu) &=& \gamma \sigma(\tilde{T},\tilde{\mu}) + 
\gamma\frac{\tilde{T}\tilde{P}_{\tilde{\mu}}}{\tilde{T}\tilde{P}_{\tilde{T}}+\tilde{\mu}\tilde{P}_{\tilde{\mu}}}\,\chi(\tilde{T},\tilde{\mu})
\eea
where $\tilde{P}_{\tilde{\mu}} \equiv \partial_{\tilde{\mu}}\tilde{P}$ and similarly for other derivatives.
In solving the equations we have introduced the integration constants $\eta,\zeta,\chi,\sigma$ above, which are each arbitrary functions of $\tilde{T},\tilde{\mu}$. We shall see that these integration constants serve as the only remaining transport coefficients for a Lorentz invariant theory. The remaining $29-6$ transport coefficients are given as follows, where we have omitted functional dependence detailed above for brevity.
\be
\begin{array}{l | l | l}
\gamma_1 = \frac{\gamma^5}{2 c^4} \zeta + \frac{\gamma^5(c^2(d-2) + d v^2)}{2 d c^6}\eta &
\gamma_2 = \frac{\gamma^5 v^2}{2 c^4} \zeta + \frac{\gamma^5(c^2 d - v^2)}{d c^4}\eta &
\gamma_3 = \frac{\gamma^3}{c^2} \zeta - \frac{2\gamma^3}{d c^2}\eta\\
\gamma_4 = 0 &
\gamma_5 = -\frac{\gamma^3}{2c^2} \eta &
\gamma_6 = -\frac{\gamma^3}{c^2} \eta \\
\gamma_7 = \frac{\gamma^3}{c^2} \eta &
\gamma_8 = \frac{\gamma^3}{2c^2} \eta &
\gamma_9 = \frac{\gamma^5}{2c^2} \zeta + \frac{(d-1)\gamma^5}{d c^2}\eta\\
\gamma_{10} = \frac{\gamma^5}{2 c^2} \zeta + \frac{\gamma^5(c^2(d-2) + d v^2)}{2 d c^4}\eta &
\gamma_{11} = \frac{\gamma^3}{c^2} \zeta - \frac{2\gamma^3}{d c^2}\eta &
\gamma_{12} = 0\\
\gamma_{13} = \frac{\gamma^3}{c^2} \eta &
\gamma_{14} = 0 &
\gamma_{15} = -\frac{\gamma^3}{2c^2} \zeta + \frac{\gamma^3}{d c^2}\eta\\
\gamma_{16} = -\frac{\gamma^3}{2c^2} \zeta + \frac{\gamma^3}{d c^2}\eta &
\gamma_{17} = 0 &
\gamma_{18} = -\frac{\tilde{T}\gamma^4}{2c^4}\chi + \frac{\gamma^4 }{2c^4} \tilde{Q}\left(\sigma+\frac{\tilde{T}\tilde{P}_{\tilde{\mu}}}{\tilde{T}\tilde{P}_{\tilde{T}}+\tilde{\mu}\tilde{P}_{\tilde{\mu}}}\chi\right) \\
\gamma_{19} = -\frac{\tilde{T}\gamma^4}{2c^2}\chi + \frac{\gamma^4 v^2}{2c^4} \tilde{Q}\left(\sigma+\frac{\tilde{T}\tilde{P}_{\tilde{\mu}}}{\tilde{T}\tilde{P}_{\tilde{T}}+\tilde{\mu}\tilde{P}_{\tilde{\mu}}}\chi\right) &
\gamma_{20} = \frac{\gamma^2 }{c^2} \tilde{Q}\left(\sigma+\frac{\tilde{T}\tilde{P}_{\tilde{\mu}}}{\tilde{T}\tilde{P}_{\tilde{T}}+\tilde{\mu}\tilde{P}_{\tilde{\mu}}} \chi\right) &
\gamma_{21} = 0\\
\gamma_{22} = 0&
\gamma_{23} = -\frac{\gamma^2}{c^2} \tilde{T}\chi &
\end{array}
\ee
where we have defined the following combination of thermodynamic quantities
\be
\tilde{Q}\equiv \frac{-\tilde{P}_{\tilde{\mu}}(\tilde{T}\tilde{P}_{\tilde{T}\tilde{T}} + \tilde{\mu}\tilde{P}_{\tilde{T}\tilde{\mu}}) + \tilde{P}_{\tilde{T}} (\tilde{T}\tilde{P}_{\tilde{T}\tilde{\mu}} + \tilde{\mu}\tilde{P}_{\tilde{\mu}\tilde{\mu}} )}{\tilde{T}(\tilde{P}_{\tilde{T}\tilde{\mu}}^2 - \tilde{P}_{\tilde{\mu}\tilde{\mu}}\tilde{P}_{\tilde{T}\tilde{T}})}.
\ee
Indeed, once the above relations are imposed, the first order constitutive relations reduce to the familiar form of relativistic hydrodynamics, viz.,
\bea
\Pi^\mu_{~\nu} &=& -\eta \Delta^{\mu\alpha} \Delta_\nu^{~\beta} \sigma_{\alpha\beta} - \zeta \Delta^\mu_{~\nu} \partial\cdot u, \quad \text{with}\quad\sigma_{\mu\nu} \equiv  \partial_\mu u_\nu + \partial_\nu u_\mu - \frac{2}{d} \eta_{\mu\nu}  \partial\cdot u,\\
\Pi^\mu &=& -\sigma \tilde{T} \Delta^{\mu\nu} \partial_\nu\left(\frac{\tilde{\mu}}{\tilde{T}}\right) + \chi \Delta^{\mu\nu} \partial_\nu \tilde{T}.
\eea
Imposing Lorentz invariance has thus reduced the number of first order transport coefficients from 29 to 4: the shear viscosity $\eta$, bulk viscosity $\zeta$, conductivity $\sigma$ and $\chi$. To reiterate, each of these are arbitrary functions of $\tilde{T},\tilde{\mu}$ in the usual way.

\subsubsection{Entropy current}
With Lorentz invariance imposed, the divergence of the canonical part of the entropy current is given by
\bea
\partial_\mu S_{\rm can}^\mu &=& - \Pi^\mu{}_\nu\partial_\mu \frac{v^\nu}{T} - \Pi^\mu \partial_\mu \frac{\mu}{T},\\
&=& - \Pi^\mu{}_\nu\partial_\mu \frac{u^\nu}{\tilde{T}} - \Pi^\mu \partial_\mu \frac{\tilde{\mu}}{\tilde{T}}.
\eea
Imposing that $S^\mu$ transform as a Lorentz vector, together with the constraints that $\partial_\mu S^\mu$ contains no explicit second derivative terms constrains the coefficients in the non-canonical entropy current \eqref{Snon0}, \eqref{Snoni} to take the form
\be
\begin{array}{llllll}
\tilde{c}_{1}=0, & \tilde{c}_{2}=0, & \tilde{c}_{3}=0, & \tilde{c}_{4}=\alpha/T^2, & \tilde{c}_{5}=0, &\tilde{c}_{6}=0,\\
\tilde{c}_{7}=-\alpha/T^2, & \tilde{c}_{8}=-\alpha/T^2, & \tilde{c}_{9}=0, & \tilde{c}_{10}=0, & \tilde{c}_{11}=0, & \tilde{c}_{12}=\alpha/T^2
\end{array}
\ee
where $\alpha$ is an unconstrained function of $\tilde{T}, \tilde{\mu}$. The non-canonical entropy current then takes the form
\be
S_{\rm non}^\mu = \alpha(\tilde{T}, \tilde{\mu})\left(u^\mu \partial\cdot u -  u^\sigma \partial_\sigma u^\mu\right),
\ee
and thus an ambiguity has appeared. However, as was shown in \cite{Bhattacharya:2011tra}, placing the theory on a curved background provides additional constraints. In particular new terms $\propto \alpha R_{\mu\nu}u^\mu u^\nu$ appear in $\partial_\mu S^\mu$, and since depending on the chosen background curvature this term can take any sign, it forces $\alpha = 0$. Adopting this result the entropy current is given entirely by the canonical piece. 

Without loss of generality we now extract positivity constraints by checking the quadratic form for fluctuations around $v^i=0$. We find,
\be
\tilde{T}\partial_\mu S^\mu = \frac{1}{2}\eta\sigma_{ij} \sigma^{ij} + \zeta(\partial_i v^i)^2 + \sigma \left(\partial_i \tilde{\mu} - \left(\frac{\tilde{\mu}}{\tilde{T}} + \frac{\chi}{2\sigma}\right) \partial_i \tilde{T}\right)^2 - \frac{\chi^2}{4 \sigma} (\partial_i\tilde{T})^2,
\ee
and hence imposing $\partial_\mu S^\mu \geq 0$ enforces
\be
\eta \geq 0,\quad \zeta \geq 0,\quad \sigma \geq 0,\quad \chi = 0.
\ee

\subsection{Galilean boosts \label{sec:galilei}}
Another special case of interest are theories with non-relativistic boost symmetry, extending the symmetry algebra of the system $H, P_i, J_{ij}, M$ for time translations, spatial translations, spatial rotations, and U$(1)$ respectively, to include a boost generator $K_i$. A particularly simple example can be reached by starting with a relativistic theory and sending $c\to\infty$, corresponding to a group contraction from ISO$(1,d)\times$U$(1)$. The resulting theory is invariant under so-called \emph{massless} Galilean boosts, for which, most notably, 
\be
[K_i, P_j] = 0.
\ee
This algebra will be the present focus of this section, which can be reached by contracting the results we have obtained in section \ref{sec:lorentz}. 
It is important to note that since we also have particle number / charge conservation, this algebra can be centrally extended to the Bargmann algebra \cite{Bargmann:1954gh} (see also \cite{Andringa:2010it}), namely,
\be
[K_i, P_j] = i M \delta_{ij},
\ee
where $M$, as the U$(1)$ generator, is the centre. This is part of a rich vein of research into non-relativistic boost-invariant hydrodynamics and couplings to gravity via the Newton-Cartan formalism. We will not consider the centrally extended case in this paper, but simply refer the interested reader to important papers in this hydrodynamic context \cite{Jensen:2014ama,Jensen:2014wha,Geracie:2015xfa,deBoer:2017ing,deBoer:2017abi}.

The limit $c\to \infty$ for the relativistic boost \eqref{LorentzBoostCoords}, \eqref{LorentzBoostVel}, \eqref{LorentzBoostTM} results in the Galilean boost with parameter $u^i = c\beta^i$,
\bea
t\to t, \quad x^i \to x^i + u^i t,\quad v^i \to v^i +  u^i, \quad T \to T,\quad \mu \to \mu.
\eea
In particular the Lorentz factor $\gamma \to 1$ as $c\to\infty$, and $\tilde{T} = T, \tilde{\mu} = \mu$, with the equation of state $P = \tilde{P}(T,\mu)$. From this we conclude,
\be
\rho = 0, \quad \tilde{\cal E} = {\cal E},\quad \tilde{n} = n,
\ee
and an ideal stress tensor (see also \cite{deBoer:2017ing})
\bea
&T^{(0)}{}^0{}_0 = -{\cal E}, \quad T^{(0)}{}^i{}_0 = -({\cal E}+P)v^i ,\quad T^{(0)}{}^0{}_j = 0, \quad T^{(0)}{}^i{}_j = P \delta^i_j\\
&J^{(0)}{}^0 = n, \quad J^{(0)}{}^i = n v^i
\eea
where $n,{\cal E}, P$ are each functions of $T,\mu$ only.

At first hydrodynamic order, recall that in the relativistic case we have 3 transport coefficients remaining after analysis of the entropy current: $\eta, \zeta, \sigma$, each a function of $\tilde{T},\tilde{\mu}$. In the $c\to\infty$ limit, each of these are functions of $T,\mu$. We can choose how each of these transport coefficients scale with $c$ so that they provide finite contributions to the constitutive relations as $c\to\infty$. This is achieved without any additional rescaling, 
\be
\eta = O(c)^0,\quad \zeta = O(c)^0,\quad \sigma = O(c)^0 \quad \text{as} \quad c\to\infty
\ee
then the 29 transport coefficients of the general theory take on the following values,
\be
\bar{\eta} = \eta,\quad \bar{\zeta} = \zeta, \quad \bar{\sigma} = \sigma, \quad \text{others}=0,
\ee
and the first order constitutive relations \eqref{constitutive} become
\bea
\Pi^{0}\!_{0} &=& 0,\\
\Pi^{0}\!_{j} &=& 0,\\
\Pi^{i}\!_{0} &=& \frac{\eta}{2} \partial_i v^2 +\eta v^k\partial_{k} v^i +\left(\zeta - \frac{2 \eta}{d}\right)  v^i\partial_k v^k,\\
\Pi^{i}\!_{j} &=&  -\zeta \delta^{i}_{j}\partial_k v^k - \eta \sigma_{ij}, \\
\Pi^0 &=& 0,\\
\Pi^i &=& -  \sigma T\partial_i \frac{\mu}{T},
\eea
where $\eta,\zeta,\sigma$ are arbitrary non-negative functions of $T,\mu$.

\subsection{Lifshitz scale invariance\label{sec:lifshitz}}
In this section we compute the form of the 29 transport coefficients for the $\mathbb{R}_t\times$ISO$(d)\times$U$(1)$ theory in the case where we further restrict to invariance under the inhomogeneous scale transformation
\be
t\to \lambda^z t, \quad x^i\to\lambda x^i, \label{LifshitzScaleTransformation}
\ee
for some arbitrary dynamical critical exponent $z$. At first sight, such invariance is merely imposed by restricting all terms in the constitutive relations to have the correct scaling weights. However, the hydrodynamic theory is treated here as an effective description of an underlying microscopic theory with a Ward identity for scale transformations. This Ward identity imposes further constraints which causes transport coefficients, or linear combinations thereof, to vanish. In the case of scale transformations this point was made clear in \cite{Son:2005tj}, where it was shown that conformal invariance leads to vanishing of bulk viscosity. We direct the interested reader to other results on Lifshitz invariant hydrodynamics \cite{Hoyos:2013eza, Hoyos:2013qna, Hoyos:2013cba, Hoyos:2015lra, Taylor:2015glc,Hartong:2016nyx, Fernandez:2019vcr}.

Let us begin with a discussion of the scaling weights. It is convenient to denote a quantity that scales as $\lambda^{-w}$ to have scaling weight $w$.
In other words, the scaling weights of $t$ and $x^i$ as presented above, are $w_t = -z$ and $w_{x^i}=-1$ respectively, whilst the scaling weights of the hydrodynamic variables ($T,v^i,\mu$) are, 
\be
 w_T = z, \quad w_{v^i} = z-1,\quad w_\mu = z.
\ee
In particular, for a transport coefficient $\gamma_I(T,v^2,\mu)$ with scaling weight $w_I$, it must be an arbitrary function of the scaling invariant combinations $v^2/T^{\frac{2(z-1)}{z}}$ and $\mu/T$, together with an overall factor of $T$ to make up its weight, i.e.
\be
\gamma_I(T,v^2,\mu) = T^{\frac{w_I}{z}} \hat{\gamma}_I\left(\frac{v^2}{T^{\frac{2(z-1)}{z}}}, \frac{\mu}{T}\right). \label{LifFunctionalForm}
\ee
This is a severe restriction on the functional form of the 29 transport coefficients, albeit not a reduction in their number. The scaling weights for the transport coefficients are as follows,
\be
\begin{array}{l | l | l | l | l}
w_{\bar{\eta}} = d & w_{\bar{\zeta}} = d & w_{\bar{\alpha}} = d-z & w_{\bar{\gamma}} = d - z & w_{\bar{\sigma}} = d-2 \\
w_{\gamma_1} = d + 4-4z & w_{\gamma_{12}} = d & w_{\gamma_{14}} = d & w_{\gamma_{17}} = d & w_{\gamma_{18}} = d + 2- 3z\\
w_{\gamma_{19}} = d-z & w_{\gamma_{20}} = d-z & w_{\gamma_{21}} = d-z & w_{\gamma_{22}} = d-z & w_{\gamma_{23}} = d-z
\end{array}
\ee
while the weights for the remaining coefficients ($\bar{\pi}, \gamma_2, \gamma_3, \gamma_4, \gamma_5, \gamma_6, \gamma_7, \gamma_8, \gamma_9, \gamma_{10}, \gamma_{11}, \gamma_{13}, \gamma_{15}, \gamma_{16}$) are each $d+2-2z$.

We now turn to the Ward identity, for which we impose the following relation
\be
z T^0_{~0} + T^i_{~i} = 0. \label{LifWard}
\ee
At ideal hydrodynamic order \eqref{LifWard} corresponds to an appropriate restriction of the equation of state. Namely,
\be
d P - z {\cal E} + v^2 \rho = 0
\ee
which implies the following PDE for the equation of state, through standard thermodynamic relations,
\be
\left(z T\partial_T  + 2(z-1)v^2 \partial_{v^2}  + z\mu \partial_\mu - (d+z)\right) P(T,v^2,\mu) = 0.\label{LifEOSPDE}
\ee
The PDE \eqref{LifEOSPDE} has the general solution
\be
P = T^{\frac{d+z}{z}}\hat{P}\left(\frac{v^2}{T^{\frac{2(z-1)}{z}}}, \frac{\mu}{T}\right),
\ee
which is of course the expected functional form for the equation of state of a Lifshitz invariant system given the scaling weight of $P$, i.e. it is of the form \eqref{LifFunctionalForm} with $w_P = d+z$. If we furthermore impose \eqref{LifWard} at first hydrodynamic order, there are four constraints, as the coefficients of the four possible scalar terms $\left\{v^k \partial_k v^2, v^k \partial_k\frac{\mu}{T}, \partial_i v^i, \partial_t v^2\right\}$. These constraints can be expressed as follows,
\bea
\bar{\zeta} &=& \frac{(z \gamma_3 -\gamma_{11}) v^2}{d},\\
\gamma_{15} &=& \frac{-2z \gamma_2+ 2\gamma_8 + 2 \gamma_9 + \gamma_{13}}{2(d-1)},\\
\gamma_{16} &=& \frac{z \bar{\pi} + 2z \gamma_1 v^2- \gamma_7 -2\gamma_{10}}{2(d-1)},\\
\gamma_{17} &=& \frac{T z(\bar{\alpha}+\bar{\gamma}) - z \gamma_4 v^2 + \gamma_{12} + \gamma_{14}}{d-1}.
\eea

To summarise this section, imposing Lifshitz scaling weights and the Ward identity \eqref{LifWard} reduces the number of transport coefficients from 29 to 25, and moreover places stringent constraints on the functional form of all of them \eqref{LifFunctionalForm}.

 \section{Discussion}
 In this paper we have constructed the first-order hydrodynamic theory describing a fluid in the presence of a preferred reference frame, which possesses no boost invariance, neither Galilean nor Lorentzian. In this frame the theory is rotationally invariant and ISO$(d)$ acts naturally. If we nevertheless boost to a reference frame with boost parameter $\beta^i$, the resulting theory will contain explicit dependence on $\beta^i$ and will no longer be rotationally invariant. Of course ISO$(d)$ is still preserved, however it acts in a less natural way. The symmetry is realised by boosting back into the preferred frame, where rotation invariance is manifest, and then boosting back to the finite $\beta^i$ frame.

A consequence of this relaxed symmetry group is the appearance of many new transport coefficients. In principle each of these transport coefficients is associated to a distinct physical effect which can be accessed by considering general fluid flows with respect to the preferred reference frame. Some of these are accessible in hydrodynamic modes. Indeed, in section \ref{sec:shear} we computed the shear diffusion mode, which grants independent access to the combinations of transport coefficients $\bar{\eta}$ and $\bar{\pi} + \gamma_6 - \gamma_7 + \gamma_{13}$ through varying the angle of the mode with respect to the background fluid flow. In addition there may be coefficients that are only accessible through nonlinear considerations. We leave a more comprehensive study of the physical effects to a future publication.

We also considered the constraints resulting from imposing the positivity of entropy production for all possible fluid configurations. We constructed the general entropy current to first order in derivatives, and found all constraints that reduce its divergence to a quadratic form.\footnote{This quadratic form is available explicitly in an accompanying notebook.} By restricting this quadratic form to shearing perturbations around a general uniform flow we extracted two very simple positivity constraints, \eqref{shearentropy}, which coincide with the linear stability requirements for the shear diffusion hydrodynamic mode. Constraints may also more easily be extracted in special cases, such as those enjoying Lorentz boost invariance. 

We also enumerated all independent linear combinations of transport coefficients that are non-dissipative, i.e. that do not contribute to $\partial_\mu S^\mu$. We counted 9 such combinations, listed in \eqref{nondissB}-\eqref{nondissE}. Understanding the relation between dissipative coefficients, non-dissipative coefficients, requirements of Onsager reciprocity and the constraints arising from hydrostatic partition functions \cite{Banerjee:2012iz,Jensen:2012jh}, is an interesting problem that deserves further study.

 Finally, we note that our constitutive relations contain only parity-invariant terms. Clearly it would be interesting to extend our analysis to include parity non-invariant effects.

\section*{Acknowledgments}
We thank Dmitry Abanin, Michele Fillipone, Kristan Jensen and Karsten Kruse for discussions. We would also like to thank Jelle Hartong and Watse Sybesma for correspondence. This work has been supported by the Fonds National Suisse de la Recherche Scientifique (Schweizerischer Nationalfonds zur F\"orderung der wissenschaftlichen Forschung) through Project Grants 200021$\_$162796 and 200020$\_$182513 as well as the NCCR 51NF40-141869 “The Mathematics of Physics” (SwissMAP). BW is supported by a Royal Society University Research Fellowship.

\appendix

\bibliographystyle{utphys}
\bibliography{hydrorefs}{}

\end{spacing}
\end{document}